\documentclass[conference]{IEEEtran}
\IEEEoverridecommandlockouts
\usepackage{cite}
\usepackage{amsmath,amssymb,amsfonts}
\usepackage{graphicx}
\usepackage{microtype}

\usepackage{amsmath,epsfig}
\usepackage{supertabular}
\usepackage{wrapfig}
\usepackage{multirow}
\usepackage{array,hhline}
\usepackage{graphicx}
\usepackage{framed}
\usepackage{setspace}
\usepackage{algpseudocode,listings}
\usepackage{xcolor}
\usepackage{caption}
\usepackage{enumitem}
\usepackage{supertabular}
\usepackage{colortbl}
\usepackage[utf8]{inputenc} % allow utf-8 input
\usepackage[T1]{fontenc}    % use 8-bit T1 fonts
\usepackage{microtype}      % microtypography
\usepackage{xcolor}         % colors
\usepackage{xspace}
\usepackage{url}
\usepackage{amssymb}
\usepackage{mathtools}
\usepackage{booktabs}
\usepackage{marvosym}

\definecolor{Light}{rgb}{0.99, 0.92, 0.95}
\definecolor{deemph}{gray}{0.6}

\newcommand{\up}[1]{\textcolor{upcolor}{$\uparrow$ #1}}
\definecolor{textpurple}{RGB}{135,89,201}
\definecolor{upcolor}{RGB}{57,182,74}
\newcommand{\down}[1]{\textcolor{red}{$\downarrow$ #1}}
\newcommand{\pub}[1]{\scriptsize\textcolor{deemph}{[#1]}}

\definecolor{codegreen}{rgb}{0,0.6,0}
\definecolor{codegray}{rgb}{0.5,0.5,0.5}
\definecolor{codepurple}{rgb}{0.58,0,0.82}
\definecolor{backcolour}{rgb}{1.0,1.0,1.0}
\lstdefinestyle{mystyle}{
    backgroundcolor=\color{backcolour},
    commentstyle=\color{codegreen},
    keywordstyle=\color{magenta},
    numberstyle=\tiny\color{codegray},
    stringstyle=\color{codepurple},
    basicstyle=\ttfamily\footnotesize\fontfamily{pcr}\selectfont,
    breakatwhitespace=false,
    breaklines=true,
    captionpos=b,
    keepspaces=true,
    showspaces=false,
    showstringspaces=false,
    showtabs=false,
    tabsize=2
}

\def\BibTeX{{\rm B\kern-.05em{\sc i\kern-.025em b}\kern-.08em
    T\kern-.1667em\lower.7ex\hbox{E}\kern-.125emX}}
\begin{document}

\title{A Multimodal Transformer for Live Streaming Highlight Prediction\\
% {\footnotesize \textsuperscript{*}Note: Sub-titles are not captured in Xplore and
% should not be used}
}

% This work was supported partially by the National Natural Science Foundations of China (Grants No.62376267, 62076242) and the innoHK project.

% \author{\IEEEauthorblockN{1\textsuperscript{st} Given Name Surname}
% \IEEEauthorblockA{\textit{dept. name of organization (of Aff.)} \\
% \textit{name of organization (of Aff.)}\\
% City, Country \\
% email address or ORCID}
% \and
% \IEEEauthorblockN{2\textsuperscript{nd} Given Name Surname}
% \IEEEauthorblockA{\textit{dept. name of organization (of Aff.)} \\
% \textit{name of organization (of Aff.)}\\
% City, Country \\
% email address or ORCID}
% \and
% \IEEEauthorblockN{3\textsuperscript{rd} Given Name Surname}
% \IEEEauthorblockA{\textit{dept. name of organization (of Aff.)} \\
% \textit{name of organization (of Aff.)}\\
% City, Country \\
% email address or ORCID}
% \and
% \IEEEauthorblockN{4\textsuperscript{th} Given Name Surname}
% \IEEEauthorblockA{\textit{dept. name of organization (of Aff.)} \\
% \textit{name of organization (of Aff.)}\\
% City, Country \\
% email address or ORCID}
% \and
% \IEEEauthorblockN{5\textsuperscript{th} Given Name Surname}
% \IEEEauthorblockA{\textit{dept. name of organization (of Aff.)} \\
% \textit{name of organization (of Aff.)}\\
% City, Country \\
% email address or ORCID}
% \and
% \IEEEauthorblockN{6\textsuperscript{th} Given Name Surname}
% \IEEEauthorblockA{\textit{dept. name of organization (of Aff.)} \\
% \textit{name of organization (of Aff.)}\\
% City, Country \\
% email address or ORCID}
% }
\author{\IEEEauthorblockN{Jiaxin Deng$^{1,2,}$\IEEEauthorrefmark{1},
Shiyao Wang$^{3,}$\IEEEauthorrefmark{1},
Dong Shen$^{3}$, 
Liqin Zhao$^{3}$, 
Fan Yang$^{3}$, 
Guorui Zhou$^{3}$, and
Gaofeng Meng$^{1,2,4,}$\textsuperscript{\Letter}}
\IEEEauthorblockA{$^{1}$State Key Laboratory of Multimodal Artifcial Intelligence Systems, \\ Institute of Automation, Chinese Academy of Sciences, Beijing, China}
\IEEEauthorblockA{$^{2}$School of Artificial Intelligence, University of Chinese Academy of Sciences, Beijing, China}
\IEEEauthorblockA{$^{3}$KuaiShou Inc., Beijing, China}
\IEEEauthorblockA{$^{4}$CAIR, HK Institute of Science and Innovation, Chinese Academy of Sciences, Hong Kong, China}

dengjiaxin2022@ia.ac.cn,\{wangshiyao08,shendong,zhaoliqin,yangfan,zhouguorui\}@kuaishou.com,gfmeng@nlpr.ia.ac.cn
% <-this % stops an unwanted space
\thanks{\IEEEauthorrefmark{1}Equal contribution. \textsuperscript{\Letter}Corresponding author.}
}

% \thanks{This work was supported partially by Kuaishou through Kuaishou Research Intern Program, the National Natural Science Foundations of China (Grants No.62376267, 62076242) and the innoHK project}

\maketitle

\begin{abstract}
Recently, live streaming platforms have gained immense popularity. Traditional video highlight detection mainly focuses on visual features and utilizes both past and future content for prediction. However, live streaming requires models to infer without future frames and process complex multimodal interactions, including images, audio and text comments. To address these issues, we propose a multimodal transformer that incorporates historical look-back windows. We introduce a novel Modality Temporal Alignment Module to handle the temporal shift of cross-modal signals. Additionally, using existing datasets with limited manual annotations is insufficient for live streaming whose topics are constantly updated and changed. Therefore, we propose a novel Border-aware Pairwise Loss to learn from a large-scale dataset and utilize user implicit feedback as a weak supervision signal. Extensive experiments show our model outperforms various strong baselines on both real-world scenarios and public datasets. And we will release our dataset and code to better assess this topic.
\end{abstract}

\begin{IEEEkeywords}
Multimodal Transformer, Modality Temporal Alignment, Border-aware Pairwise Loss, Live Streaming Highlight Prediction
\end{IEEEkeywords}

\section{Introduction}
\label{sec:intro}
\begin{figure}
\centering
\includegraphics[width=.49\textwidth]{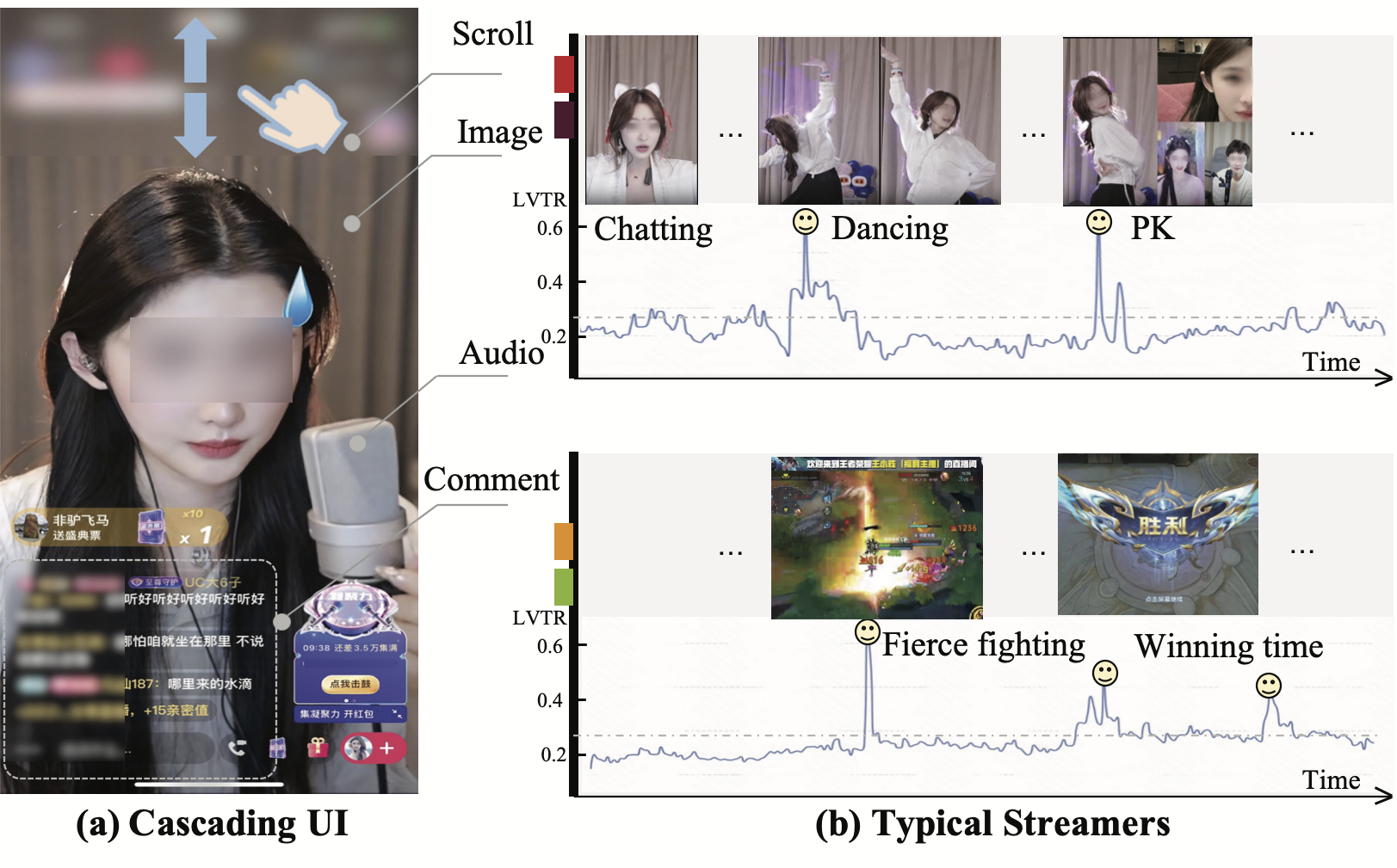}
\caption{\textbf{The live streaming platform and typical streamers. (a) Cascading UI of live streaming. (b) Highlight moments of two typical streamers.}}
\label{fig1}
\vspace{-0.8cm}
\end{figure}

Live streaming platforms represent a new type of online interaction and have experienced rapid growth in recent years. This new form of interaction and entertainment has motivated researchers to study emerging issues such as gift-sending mechanisms, E-Commerce events and other practices. As shown in Figure \ref{fig1} (a), live streaming contains complex multimodal interactions, including images, audio and text comments. And the host's streaming content may undergo dramatic topic shifts affected by the interactive comments of audiences. So an accurate live streaming understanding algorithm, which is capable of fully utilizing multimodal information during broadcasting has become paramount.

Although plenty of methods \cite{garcia2018phd,rochan2020adaptive,bhattacharya2022show,wang2022pac} have proposed to process frames or text in videos, the application and research in live streaming still suffer from lots of difficulties. First, unlike videos, live streaming makes predictions only based on information available up until that moment. Besides, multimodal information in these untrimmed videos is usually misaligned. For example, the reaction of hosts and audiences can experience a time lag, so the streamer's speech and audiences' comments may be ambiguous and not sequentially aligned with the visual frames, necessitating a module to mitigate the noise caused by misalignment. Moreover, there is no large-scale public dataset for live streaming highlight detection. AntPivot \cite{zhao2022antpivot} propose a dataset called \textit{AntHighlgiht}, but it only provides 3,656 samples with only textual feature. Hence, a large-scale live streaming dataset with multimodal information is crucial to assessing this topic. 

In this paper, we propose a transformer-based network, which leverages multi-modality features for live streaming highlight prediction. First, we formulate the task as a prediction task based on historical look-back windows and the casual attention mask is proposed to avoid the information leakage from the future. Second, to alleviate the misalignment between visual and textual modality, we develop a novel \textbf{Modality Temporal Alignment Module} to address potential temporal discrepancies that may arise during live streaming events. 
Finally, we construct a large-scale live streaming dataset, named \textit{KLive}, which provides segment-level content information like visual frames, comments and ASR results. Different from previous highlight datasets which use binary labels (highlight or non-highlight frames), KLive provides dense annotations and is more precise in reflecting the users’ general preferences on live streaming content. Based on KLive, we design a novel \textbf{Border Aware Pairwise Loss} with first-order difference constraints. We find that the constraints are essential when jointly optimizing pointwise and pairwise losses to avoid collisions and model collapse. 
We perform comprehensive experiments on both the large-scale real-world live streaming dataset KLive and a public PHD dataset \cite{garcia2018phd} and achieve state-of-the-art performance. 
In summary, the main contributions made in this work are as follows:
\begin{itemize}[leftmargin=*]
\item We propose a multimodal transformer framework for highlight detection in live streaming. For alleviating cross-modality misalignment, a modality Temporal Alignment Module is further presented to tackle this challenge.
% \item A dynamic time warping (DTW) based alignment strategy is proposed to alleviate misalignment in streaming scenarios and we design a pairwise-based loss function with first-order difference constraints to exploit the contrastive information of highlight frames and no-highlight frames.
\item We provide a new dataset, called \textit{KLive}, and design a Border Aware Pairwise Loss with first-order difference constraints to exploit the contrastive information of highlight frames and no-highlight frames.
\item Extensive experiments are conducted on both KLive and public PHD dataset, and our model achieves \textit{state-of-the-art} performance. We also present ablation and visualization results to demonstrate the effectiveness.
%We also perform online A/B testing on the live streaming platform of company, achieving a 5.9\% improvement in terms of live play duration.
\end{itemize}
\begin{figure*}[h]
\centering
\includegraphics[width=0.95\textwidth]{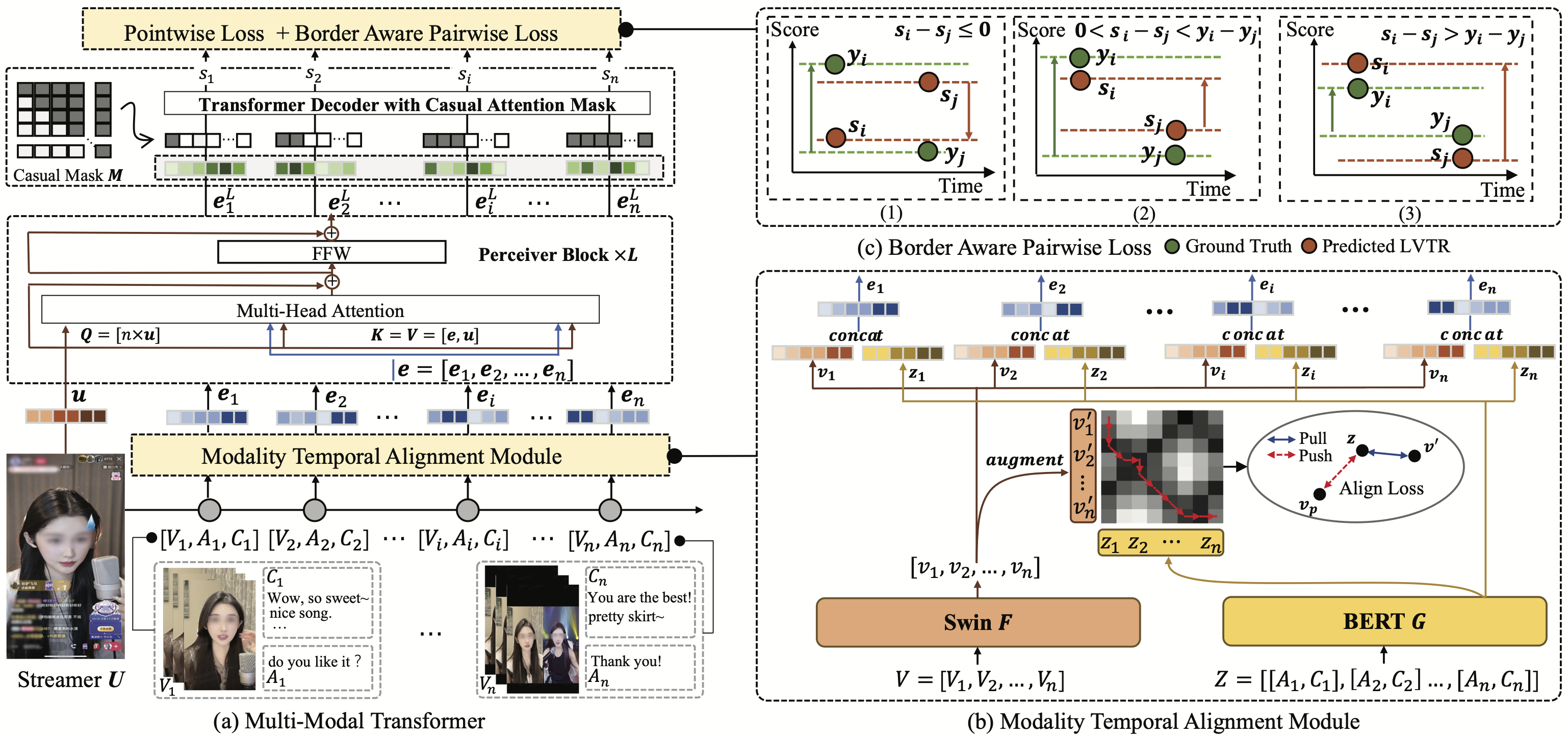}
\caption{\textbf{The framework of our method.} Part (a) shows the architecture of Perceiver Block and Casual Attention Decoder which are discussed in Sect.\ref{MultiModalTransformer}. Part (b) shows the proposed Modality Temporal Alignment Module which is discussed in Sect.\ref{DTWAlignment}. Part (c) shows the motivation of Border-aware Pairwise Loss which is discussed in Sect.\ref{borderAwarePairwiseLoss}.}
\label{fig2}
\vspace{-0.7cm}
\end{figure*}
\section{Related Work}
\subsection{Video Highlight Detection}
The task most closely related to streaming highlight prediction is Video Highlight Detection, as both aim to identify different content patterns in temporal sequences. Previous works such as dppLSTM \cite{zhang2016video} and Video2GIF \cite{feng2018extractive} attempt to exploit temporal dependence from both past and future content. However, these works fail to adapt to live streaming scenarios without future frames. Recently, several works \cite{garcia2018phd,rochan2020adaptive,chen2021pr,wang2022pac,bhattacharya2022show} try to further extract user-adaptive highlight predictions guided by annotated user history. For example, PHD-GIFs \cite{garcia2018phd} is the first personalized video highlight detection technique that also creates a large-scale dataset called PHD. \cite{wang2022pac} design a DBC module to generate user-adaptive highlight classifier. \cite{bhattacharya2022show} focus on leveraging the visual contents of both preferred clips and the target videos. We will compare with the above baselines to prove the effectiveness of our model.
%Although VHD task and stream highlight detection are quite similar, VHD only requires predicting whether frames should be highlighted or not, while stream highlight detection needs to regress the correct order of all frames without accessing the future frame, which is more challenging.

\subsection{Live Streaming Highlight Detection}
To the best of our knowledge, the concept of Live Streaming Highlight Detection (LSHD) is initially defined by \cite{zhao2022antpivot}. The task seeks to retrieve corresponding frames for highlight topics and discussions. They propose a novel \textit{Pivot Transformer} to capture temporal dependencies and integrate hierarchical semantic levels. They also construct a dataset called \textit{AntHighlight} for streaming highlight detection. However, this work mainly considers the conversation interaction.
%but fails to tackle the alignment issue among visual, speech and comment modality. 
And \textit{AntHighlight} provides 3,656 records with speech modality information while our propose dataset provides multimodal information and more dense highlight labels.

\section{Methodology}
% As illustrated in Figure \ref{fig2}, our model utilizes multi-modal features to predict the LVTR at frame level. In particular, we first define the problem of predicting LVTR for stream highlights in Section \ref{problem}. Then the Multi-Modal Transformer backbone is introduced to fuse and interact with different modalities in Section \ref{MultiModalTransformer}. Additionally, we propose the Dynamic Time Warping strategy for aligning text and visual feature in Section \ref{DTWAlignment}. Finally, border-aware pairwise loss for exploring contrasting information is presented in Section \ref{borderAwarePairwiseLoss}.
\subsection{Problem Formulation and Model Overview}
\label{problem}
\begin{spacing}{1.0}
As shown in Figure \ref{fig1} (a), the live streaming frame $\delta_i$ is shown to user with cascading UI. The streaming content will automatically play in this UI and users can choose to stop viewing current streaming room by scrolling to the last or next streaming room. Therefore, the \textit{view time} becomes an important signal to reflect the live streaming content quality, which correlates with the occurrence of highlight moments. We define the impression whose watching time is greater than a threshold as long-viewing impressions. Note an impression is defined as an instance when live streaming frame $\delta_i$ was recommended to a user \cite{nayak2023news}. Then, the Long View Through Rate (LVTR) is calculated as follows:
\end{spacing}
{
\footnotesize
\begin{equation}
LVTR=\frac{\mathrm{Number} \ \mathrm{of} \ \mathrm{Long} \ \mathrm{View}  \ \mathrm{Impressions}}{\mathrm{Number} \ \mathrm{of} \ \mathrm{the  } \ \mathrm{Total} \ \mathrm{Impressions}}
\label{lvtrdefination}
\end{equation}
}
\begin{spacing}{1.0}

From the Figure \ref{fig1} (b), we find that the high points of the LVTR curve usually correspond to the highlight moments. So we formulate the live streaming highlight prediction task by predicting the LVTR of frame $\delta_i$.
% The primary objective of live streaming highlight prediction is to determine the probability of the following content being a highlight moment. We modify this problem by predicting the LVTR of frame $\delta_i$, which can produce binary prediction results by a specified threshold.
\end{spacing}
\begin{spacing}{1.0}
We denote $M_i=\{V_i, A_i, C_i\}$ as the multi-modal tuple, where $V_i,A_i$ and $C_i$ represent the visual, audio and comments of frame $\delta_i$. Note that we use Automatic Speech Recognition (ASR) to extract information from audio. For frames $\delta_i$ at timestamp $i$, the corresponding frames feature $v_{i}$ and concated ASR and comments text feature $z_{i}$ are extracted as follows:
\end{spacing}
{\footnotesize
\begin{equation}
v_i=F\left( V_i \right) ,  z_i=G\left( \left[ A_i,C_i \right] \right) 
\label{Extraction}
\end{equation}
}
where $F(\cdot)$ is a swin and $G(\cdot)$ is BERT. We concatenate $v_{i}$ and $z_{i}$ as the final input tokens $e_{i}$. Since the highlight pattern and audience taste change over time, the model should use the multi-modal information from the $n-1$ lookahead windows $e=[e_{i-n+1},e_{i-n+2},\cdots, e_{i-1}]$ of previous frames to predict the LVTR $y_i$ of frame $\delta_i$. 

% The prediction problem can be formulated as $s_i \sim \Gamma\left( W_M, u, \delta_i \right)$, where $W_M$ is the multi-modal feature of lookahead windows, $u$ is the ID embedding of streamer $U$, $s_i$ is the predicted LVTR of frame $\delta_i$ and $\Gamma\left( \cdot \right)$ is the model to make prediction.
In this work, the multi-modal features will interact with ID embedding by the Perceiver Block and Casual Attention Decoder defined in Sect. \ref{MultiModalTransformer} and be aligned by the Modality Temporal Alignment Module in Sect. \ref{DTWAlignment}. Finally, the model will be optimized by the Pointwise and Pairwise loss defined in Sect. \ref{borderAwarePairwiseLoss}.
% \operatorname{Prob}\left(\delta_i \mid W_M, E_u \right) \sim \Gamma\left( W_M, E_u, \delta_i \right)
% \end{equation}

%公式参考mPLUG-2: A Modularized Multi-modal Foundation Model  Across Text, Image and Video
\subsection{Perceiver Block and Casual Attention}\label{MultiModalTransformer} 
% \textbf{Feature Fusion Layer}
% As depicted in Figure \ref{fig2} (d), given the historical window $W_M$ of streamer $u$, we extract multi-modal features for every timestamp, including the streaming frames $v_i$, Auto Speech Recognition (ASR) text $a_i$ from the streamer, comment $x_i$ from the audiences. The streaming frames are tokenized by the pre-trained swin \cite{liu2021Swin} $f(\cdot)$ while the ASR and comment text are tokenized with the BERT Chinese Large model $g(\cdot)$ and we only use the hidden feature of \texttt{<CLS>} token as the text embedding. Two MLP heads $proj_1(\cdot)$ and $proj_2(\cdot)$ are set to map the frames embedding and text embedding to the same dimension $d$, denoted by $S_p$ and $S_a$. 

% The above process can be formulated as follows:
% \begin{equation}
% \begin{aligned}
% {S}_{{p}} & ={proj}_1(f(\boldsymbol{v})) \\
% {S}_{{a}} & ={proj}_2\left({g}_{\texttt{<{CLS}>}}([\boldsymbol{a}, \boldsymbol{x}])\right)
% \end{aligned}
% \end{equation}
% where $\boldsymbol{v}=\left[ v_1,\cdots ,v_n \right] , \boldsymbol{a}=\left[ a_1,\cdots ,a_n \right]$ and $ \boldsymbol{x}=\left[ x_1,\cdots ,x_n \right] $.
% \subsection{Perceiver Block}\label{PerceiverBlock}
\begin{spacing}{1.0}
\textbf{Perceiver Block} We hypothesize that different streamers have distinct talents and attract different audiences.
%who are interested in specific types of highlight moments. 
%as shown in Figure \ref{fig2} (a), 
For instance, as shown in Figure \ref{fig1} (b), some audiences like dancing, while others may be attracted to the PK events between streamers. Therefore, we use separated ID embeddings $u$ for streamer $U$ to extract streamer-aware multimodal features (see Figure \ref{fig2} (a)). Note $u$ will be repeated $n$ times to match the sequence length of the multimodal features $e$. 
%So proposed Perceiver block takes the repeated ID embedding $n \times u$ of streamer $U$ and the fusion multi-modal feature $e$ as input. 
First, we initialize learned latent query $Q$ with flattened $n \times u$. Next, we concatenate the flattened $e$ and $n \times u$ at the second dimension and take $K = V = [e, n \times u]$ as key and value. 
%We linearly project the input vector to latent vectors with $d_h$ dimensions, by different linear projections. 
Then we perform standard scaled dot-product multi-head attention on the projected vectors with feed-forward network and residual connections. 
The Perceiver block is stacked for $L$ layers and the final output of Perceiver block is denoted by $e^L$.

\textbf{Transformer Docoder} We unflatten and squeeze the output of Perceiver block $e^L$ as the dimension of $\mathcal{R}^{b \times n \times d}$. Then we initialize the query, key and value of the decoder as $Q = K = V = e^L$ and we apply the scaled dot-product attention function with casual attention as follows:
\end{spacing}
{
\footnotesize
\begin{equation}
\operatorname{Attention}(Q, K, V)=\operatorname{softmax}\left(\frac{Q K^{\top}}{\sqrt{d_h}} + M \right) \cdot V
\label{atten3}
\end{equation}
}
where $M$ is the casual attention mask and $d_h$ is the hidden dimension. As shown in Figure \ref{fig2} (a), it is an $n \times n$ matrix filled with \text{-inf} and its upper triangular sub-matrices are filled with 0. By applying the casual attention mask $M$ the problem of future information leakage in the temporal dimension is avoided. 
% The pseudocode of the Decoder is shown in Algorithm \ref{alg2}.
% \begin{algorithm}[h]
% \caption{A Pytorch-style Pseudocode for Decoder.}
% \label{alg2}
% \begin{lstlisting}[language=python]
% def decoder(
%     x, # The input with shape [b*n, 1, d_n]
%     num_layers, # The number of layers
% ): 
%     x = unflatten(x).squeeze() #[b*n,1,d_n] -> [b,n,d_n]
%     for i in range(num_layers):
%         # Attention
%         x = x + attention_with_mask_i(q=x, k=x, v=x))
%         # Feed forward with residual connection
%         x = x + ffw_i(x)
%     return x
% \end{lstlisting}
% \end{algorithm}
The output from the final attention layer is then fed into a fully connected layer, followed by a sigmoid transformation to produce the scalar prediction of LVTR $s$.

\subsection{Modality Temporal Alignment Module (MTAM)}\label{DTWAlignment}
\begin{spacing}{1.0}
The motivation behind for alignment is to address potential temporal discrepancies that may arise during live streaming events. For example, the streamer may describe the plan before taking action, or explain detailed information after action. Additionally, the comments from the audiences may experience some time lag, which further exacerbates the misalignment issue. Therefore, it is essential to train text and visual encoders that can handle misalignment to alleviate that problem. Inspired by previous works \cite{ko2022video}, in this section we present our contrastive learning-based framework for visual and text sequence alignment.
\end{spacing}

\begin{spacing}{1.0}
% we assume that when timestamps $i$ in the text sequence and $j$ in the visual sequence represent the same pattern, then $S_a^{i}$ and $S_p^{j}$ should have a semantic similarity. 
% In order to train a network that can minimize the distance between the text sequence $S_a$ and the visual sequence $S_p$, we utilize Dynamic Time Warping (DTW) to calculate the minimum cumulative matching costs over units as the sequence distance to measure the similarity. 
In order to align $z$ and $v$ with optimal matching correspondence while considering the constraint of temporal order, we utilize the Dynamic Time Warping (DTW) algorithm by calculating the minimum cumulative matching cost between two sequences. Specially, we first compute a pairwise distance matrix $D\left( z,v \right) \coloneqq \left[ \mathrm{sim} \left( z_{i},v_{j} \right) \right] _{ij}\in \mathcal{R} ^{n\times n}$ with a distance measure $\mathrm{sim}(\cdot)$. In our work, we apply the cosine similarity as $\mathrm{sim}(\cdot)$. Then, we employ dynamic programming and sets a matrix $H \in \mathcal{R}^{n \times n}$ to record the minimum cumulative cost between $z_i$ and $v_j$:
\end{spacing}
% \footnotesize
{
\footnotesize
\begin{equation}
H_{i,j} =D_{i,j} +\min \left\{ H_{i-1,j-1} ,H_{i-1,j} ,H_{i,j - 1} \right\} 
\label{dtwdp}
\end{equation}
}
where $1\le i,j\le n$. Then, the distance $d_{\left\{z, v\right\}}$ between sequences $z$ and $v$ is set to the last element of matrix $C$:
{\footnotesize
\begin{equation}
d_{\left\{z, v\right\}}=H_{n,n}
\label{dtwdpdis}
\end{equation}
}
\begin{spacing}{1.0}
However, the standard DTW cannot solve the non-sequential alignments since it allows only three movement directions $\left \{\downarrow,\searrow,\rightarrow\right \} $. To tackle this problem, we propose a novel augmentation method to partially shuffle the original video sequence. Let $\omega$ denote all possible time index pair combinations retrieved from $v$, we formulate the target distribution as follows:
\end{spacing}
{\footnotesize
\begin{equation} \label{dtwdpdis1111}
\begin{aligned}
p^{video}=\mathrm{softmax} \left( \frac{D\left( z,v \right) _{ij}}{\gamma} \right) , \left( i,j \right) \in \omega 
\end{aligned}
\end{equation}
}
\begin{spacing}{1.0}
where $\gamma$ is a temperature parameter. The proposed target distribution is more likely to generate a time index pair that has the most similarity between $z$ and $v$. Then, the index pair $(i,j)\sim P^{video}$ is sampled from the distribution $P^{video}$ defined in Equation \ref{dtwdpdis1111} and we swap the corresponding values of $v$ to generate a new positive sample $v^{'}$. In this way, the positive sample $v'$ is more likely to have a more optimal temporal alignment order with $z$. Because compared to the original sequence $v$, there may exist no sequential alignment instances between original $v$ and $z$. We hypothesize that the positive pair $\left\{z, v^{'}\right\}$ should be similar, while the negative pair $\left\{z, {v}_p\right\}$ should be dissimilar, where ${v}_p$ is directly shuffled from $v$. Consequently, we can formulate the training objective as minimizing the well-known InfoNCE loss:
\end{spacing}
{\footnotesize
\begin{equation}
\begin{aligned}
\mathcal{L}_{align}= -\log \frac{\exp \left( d_{\left\{ z,v^{'} \right\}}/\tau \right)}{\exp \left( d_{\left\{ z,v^{'} \right\}}/\tau \right) +\sum_{v^i_p\in \omega}^N{\exp}\left( d_{\left\{ z,v^i_p \right\}}/\tau \right)}
\end{aligned}
\end{equation}
}
where $N$ is the number of negative samples directly shuffled from $v$. As shown in Figure \ref{fig2} (b), by minimizing the align loss $\mathcal{L} _{align}$, the visual encoder $F(\cdot)$ and text encoder $G(\cdot)$ are encouraged to learn a good representation from aligned and misaligned pairs.

\subsection{Border Aware Pairwise Loss}\label{borderAwarePairwiseLoss}
% In this section, we present some intriguing discoveries on the traditional pairwise logistic ranking loss \cite{burges2005learning} and propose modifications to the loss function by integrating the border-aware first-order difference constraint, thereby enhancing the optimization. 
% The addition of the pairwise loss aids in exploiting the underlying contrasting information present in both the highlight and no-highlight frames.
\begin{spacing}{1.0}
The objective of pointwise loss is to maximize the log-likelihood between the predicted LVTR $s$ and the actual LVTR $y$, which can be achieved using the following pointwise model to optimize the standard LogLoss \cite{liu2009learning}:
\end{spacing}
{
\footnotesize
\begin{equation}\label{PointWiseLoss}
L_{Point} = -\frac{1}{n}\sum_{i=1}^{n}\left[y_i\cdot\log\left(s_i\right) + (1-y_i)\cdot\log\left(1-s_i\right)\right]
\end{equation}
}

However, the pointwise loss defined in Equation \ref{PointWiseLoss} may fail to effectively exploit the contrastive information between the highlight and no-highlight frames. Therefore, we introduce the pairwise loss for further optimization. In this section, we present some intriguing discoveries on the traditional pairwise logistic ranking loss \cite{burges2005learning} and propose modifications to the loss function by integrating the border-aware first-order difference constraint.

\begin{figure}[h]
\centering
\includegraphics[width=.45\textwidth]{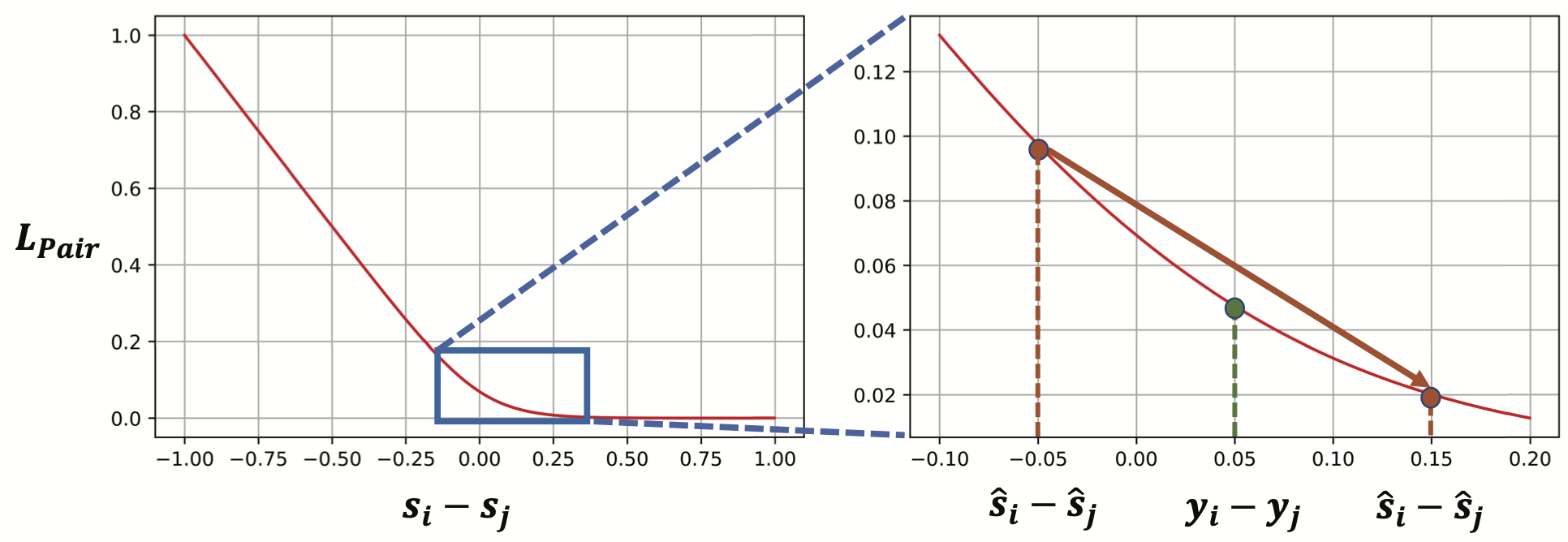}
\caption{\textbf{The change of pairwise loss function w.r.t. $s_i-s_j$ .}}
\label{fig3}
\vspace{-0.4cm}
\end{figure}
Consider the following pairwise loss function, which has no constraints:
% \footnotesize
{
\footnotesize
\begin{equation}
L_{Pair}^0=\sum_{y_i>y_j}{\log \left( 1+e^{-\sigma\left( s_i-s_j \right)} \right)}
\label{paitlsoos}
\end{equation}
}
where $y_i$ and $y_j$ are the ground truth LVTR at timestamps $i$ and $j$, $s_i$ and $s_j$ is the predicted LVTR from the model and $\sigma$ is a scale hyperparameter. Figure \ref{fig3} illustrates the changing of the loss function $L_{Pair}$ with respect to $s_i-s_j$, which reveals that minimizing $L_{Pair}$ tends to cause $s_i-s_j$ to overtake the optimal value of $y_i-y_j$. This leads to over-optimization. 

\begin{spacing}{1.0}
Based on the above findings, we propose a revised pairwise loss with a border-aware first-order difference constraint:
\end{spacing}
{\footnotesize
\begin{equation}
L_{Pair}^1=\sum_{y_i>y_j}{\log \left( 1+e^{-\sigma \left( s_i-s_j \right)} \right)},\left( y_i-y_j \right) -\left( s_i-s_j \right) \geqslant 0
\end{equation}
}
where $\left( y_i-y_j \right) -\left( s_i-s_j \right) \geqslant 0 $ denotes the border, and solely those samples that reside within the border will calculate $L_{Pair}^1$. Any samples located outside the border will result in $L_{Pair}^1$ being set to 0.

Without losing generality, the original pairwise loss function $L_{Pair}^0$ presented in Equation \ref{paitlsoos} can be divided into three distinct parts:
\begin{itemize}
\item Part 1: As shown in Figure \ref{fig2} (c1), when $s_i-s_j \leq 0$, the model's assessment of the significance between timestamp $i$ and $j$ is entirely incorrect, given that timestamp $i$ is more "highlighting" than timestamp $j$.
\item Part 2: As shown in Figure \ref{fig2} (c2), when $0 < s_i-s_j < y_i-y_j$, it implies that the model has distinguished that timestamp $i$ is more "highlighting" than timestamp $j$, but it still fails to accurately predict the difference in LVTR values between the two timestamps, and thus, it is still not optimal.
\item Part 3: As shown in Figure \ref{fig2} (c3), when $ y_i-y_j < s_i-s_j$, it indicates that the model's predicted LVTR value is too aggressive, resulting in over-optimization.
\end{itemize}

In order to verify the above scenarios, we design the following loss functions:
% \footnotesize
{\footnotesize
\begin{equation}
L_{Pair}^{2}=\sum_{y_i>y_j}{\log \left( 1+e^{-\sigma \left( s_i-s_j \right)} \right)},s_i-s_j \leq 0
\end{equation}
}
where $s_i-s_j\leq0$ means that it only optimizes on Part 1.
{\footnotesize
\begin{equation}
\begin{aligned}
L_{Pair}^{3}=\sum_{y_i>y_j}{\log \left( 1+e^{-\sigma \left( s_i-s_j \right)} \right)},y_i-y_j>s_i-s_j>0
\end{aligned}
\end{equation}
}
where $y_i-y_j>s_i-s_j>0$ means that it only optimizes on Part 2. The ablation study among  $L_{Pair}^{0}$,  $L_{Pair}^{1}$,  $L_{Pair}^{2}$ and  $L_{Pair}^{3}$ is discussed in Section \ref{exper}. In this work, we apply the $L_{Pair}^{1}$ for optimization.

\begin{spacing}{1.0}
By combining pointwise loss in section \ref{MultiModalTransformer}, align loss in section \ref{DTWAlignment} and pairwise loss, our final loss used to learn the model parameters is defined as:
\end{spacing}
% \footnotesize
{\footnotesize
\begin{equation}
\mathcal{L}=\lambda_{1}\mathcal{L}_{Point}  + \lambda_{2}\mathcal{L}_{align} +  \lambda_{3}\mathcal{L}_{Pair}^1
\label{multiloss}
\end{equation}
}
where $\lambda_{1}$, $\lambda_{2}$ and $\lambda_{3}$ are the tradeoff parameters.
\begin{table*}[htbp]
    \centering
    \caption{
    Performances of different methods on KLive and PHD dataset
    }
    \label{overall}
    % \vspace{-8pt}
    \setlength{\tabcolsep}{3.1pt}
    \scalebox{0.7}{
    \begin{tabular}{p{4.0cm}<{\raggedright}p{2.5cm}<{\centering}|p{3.5cm}<{\centering}p{3.5cm}<{\centering}p{3.5cm}<{\centering}p{3.5cm}<{\centering}|p{3.5cm}<{\centering}}
    \hline
    \multicolumn{2}{c|}{\multirow{2}{*}{\textbf{Methods}}}  &   \multicolumn{4}{c|}{\textbf{KLive} \textbf{Tau} $\boldsymbol{\tau \uparrow}$ } & \multicolumn{1}{c}{\multirow{2}{*}{\textbf{PHD mAP} $\boldsymbol{\uparrow}$}}
    \\
    % \multicolumn{2}{c}{}  &   \multicolumn{5}{c}{\textbf{Tau} $\boldsymbol{\tau \uparrow}$}  & \multicolumn{1}{c}{\multirow{2}{*}{\textbf{mAP} $\boldsymbol{\uparrow}$}}\\
    \multicolumn{2}{c|}{}                       &   $\Delta=0$&  $\Delta=0.2$&  $\Delta=0.4$ & $\Delta=0.6$ & \multicolumn{1}{c}{}
    \\ 
    \hline
    \multicolumn{1}{l}{\textbf{\textit{VHD Methods}}} \\
    \hline
    \textbf{Adaptive-H-FCSN} \cite{rochan2020adaptive} & \pub{ECCV'20} &0.5782 & 0.5707 & 0.5511 & 0.5322  &15.65 \\ %\up{0.xx\%}
    \textbf{PR-Net} \cite{chen2021pr}  & \pub{ICCV'21} & 0.5848 &  0.5818 & 0.5461 & 0.5403 &  18.66\\ %\up{0.xx\%}
    \textbf{PAC-Net} \cite{wang2022pac}  & \pub{ECCV'22} &  0.5823  & 0.5845  & 0.5537 & 0.5409 & 17.51\\ %\up{0.xx\%}
    \textbf{ShowMe} \cite{bhattacharya2022show}  & \pub{MM'22} & 0.5798   &  0.5705  & 0.5348  & 0.5407   &  16.40\\ %\up{0.xx\%}
    \hline
    \multicolumn{1}{l}{\textbf{\textit{LSHD Methods}}}  \\
    \hline
    \textbf{AntPivot} \cite{zhao2022antpivot}  & \pub{arXiv'22} & 0.5818  & 0.5809  & 0.5483  & 0.5421    & - \\ %\up{0.xx\%}
    % \textbf{SelfAttention-plain}   &  0.5868 \\
    % \textbf{SelfAttention-input}   &   0.5718 \\ %\down{1.50\%}
    % \textbf{CrossAttention-q}   &   0.5909 \\ %\up{0.41\%}
    % \textbf{CrossAttention-kv}   &  0.5780 \\  %\down{0.88\%}å
    % \textbf{CrossAttention-qkv}   &    0.5883 \\ %\up{0.15\%}
    % \hline
    \cellcolor{Light}{\textbf{KuaiHL}} &\cellcolor{Light}{\pub{Ours}} &  \cellcolor{Light}{\textbf{0.5961} } &   \cellcolor{Light}{\textbf{0.5871} } &   \cellcolor{Light}{\textbf{0.5686} } &   \cellcolor{Light}{\textbf{0.5563} } &    \cellcolor{Light}{\textbf{21.89}} \\ %\up{0.51\%}}
    \hline
    \end{tabular}}
    \vspace{-8pt}
\end{table*}

\section{Experiments}\label{exper}
In this section, we perform experiments with KLive and PHD dataset and present the experimental results and some analysis of them. Please refer to the supplementary material for more information of dataset and implementation details. Our code is available at \url{https://github.com/ICME24/KLive}.
% Specifically, our experiments aim to answer the following research questions: \textbf{RQ1}: How well does the proposed model perform on public datasets, such as video highlights? \textbf{RQ2}: How does the modality impact the model? \textbf{RQ3}: How much can the model's performance be enhanced by the proposed pairwise loss? \textbf{RQ4}: How effective is the proposed DTW alignment block in alleviating misalignment cases? 
% \textbf{RQ5}: To what extent can the proposed Perceiver Block enhance the model's performance?
\subsection{Dataset}\label{Dataset}
To comprehensively evaluate the model's performance, we show results on both KLive and public PHD dataset \cite{garcia2018phd}.

%\textbf{KLive Dataset} We construct a large-scale dataset which contains nearly 19,334 hours live streaming records from a well-known live streaming platform. We select 17,897 high-quality live rooms based on a set of predefined criteria. Each live room is then divided into multiple consecutive 30s live segments, with three pictures evenly sampled for each segment, the streamer's ASR and audiences' comments. We expose the live frame segments obtained above to users and calculate the LVTR defined in Equation \ref{lvtrdefination}. We then construct consecutive 20 streaming segments into a dataset sample, filtering out samples with a low number of watched users of the last live segment to ensure a reliable LVTR. We select 14,897 live rooms as the training set, while the remaining 3,000 live rooms as the test set. In the end, we obtained 1,436,979 and 286,510 samples for our training and test datasets. Because the objective of our method is to predict the highlight moment of future frames, we left shift the ground truth LVTR for all live segments as the final label.
\textbf{KLive Dataset} We construct a large-scale dataset that contains 17,897 high-quality live rooms (19,334 hours) from a well-known live streaming platform. Each live room is divided into multiple consecutive 30s live segments, with three pictures evenly sampled for each segment, the streamer's ASR and audiences' comments. We employ consecutive 20 segments as a sample and obtain 1,436,979 and 286,510 samples for training and test datasets. Because the objective is to predict the \textit{LVTR} of \textbf{future} frames, we left shift the ground truth for all live segments as the final label.
%% 放到 SUpplemtary

\textbf{PHD Dataset}
To verify the generality of our method, we evaluate on the publicly available video highlight detection dataset \cite{garcia2018phd} (PHD). The training set comprises of 81,056 videos, while the testing set has 7,595 videos. Please refer to supplementary material for more details.

\subsection{Evaluation Metrics}
\begin{spacing}{1.0}
For the experiments on KLive dataset, we employ rank correlation coefficients Kendall’s tau $\tau$ \cite{saquil2021multiple} to measure the correlation between our predicted LVTR $s$ and the ground truth LVTR $y$. We also report the various levels of Kendall's tau agreement for the live frames whose ground truth LVTR is greater than the threshold $\Delta$.
% It is defined as:
% \end{spacing}
% % \footnotesize
% {
% \begin{equation}
% \tau =\frac{P-Q}{\sqrt{\left( P+Q+T \right) \cdot \left( P+Q+U \right)}}
% \end{equation}
% }
% where $P$ is concordant pairs number, $Q$ is discordant pairs number, $T$ is the number of ties only in $s$, and $U$ is the number of ties only in $y$. If the same pair experiences a tie in both $s$ and $y$, it is not included in either $T$ or $U$. 
On PHD dataset, we utilize the widely adopted mean Average Precision (mAP) as a metric to evaluate the performance of our method, which is also applied in previous works \cite{bhattacharya2022show,wang2022pac,chen2021pr,rochan2020adaptive} in video highlight detection. We report the mAP on the test set and follow the way in \cite{wang2022pac} to calculate the mAP.
\end{spacing}
\subsection{Overall Performance Comparison}
%参考mPLUG-2: A Modularized Multi-modal Foundation Model  Across Text, Image and Video
Table \ref{overall} summarizes the LVTR prediction performances achieved by various methods on KLive and PHD dataset. On the KLive dataset, KuaiHL exhibits superior performance compared to all VHD and LSHD methods across various threshold levels, surpassing the LSHD method \textbf{AntPivot} which only models the text modality by +1.43\% in tau. Similarly, on PHD dataset, KuaiHL outperforms the strongest baseline \textbf{PR-Net} \cite{chen2021pr} by +3.23\%. 

\subsection{Border Aware Pairwise Loss and MTAM}
We investigate the impact of different loss functions on KLive dataset, which include $L_{Point}$, $L_{Pair}^{0}$, $L_{Pair}^{1}$, $L_{Pair}^{2}$, $L_{Pair}^{3}$, and $L_{align}$ (Note that alignment loss without the augmentation method defined in Equation \ref{dtwdpdis1111} is represented as $L_{align}^{'}$). 
% \vspace{-6pt}
\begin{table}[h]
\vspace{-3pt}
    \centering
    \caption{
    Ablation study of KuaiHL with different loss functions on KLive dataset.
    }
    \label{lossfunc}
    \vspace{-4pt}
    \setlength{\tabcolsep}{3.1pt}
    \scalebox{0.7}{
    \begin{tabular}{p{1.2cm}<{\centering}p{1.cm}<{\centering}p{1.cm}<{\centering}p{1.cm}<{\centering}p{1.cm}<{\centering}p{1.cm}<{\centering}p{1.cm}<{\centering}p{1.cm}<{\centering}|l}
    \toprule
    \textbf{Methods} & $\boldsymbol{L_{Point}}$ & $\boldsymbol{L_{Pair}^{0}}$ & $\boldsymbol{L_{Pair}^{1}}$ & $\boldsymbol{L_{Pair}^{2}}$ & $\boldsymbol{L_{Pair}^{3}}$ & $\boldsymbol{L_{align}^{'}}$ & $\boldsymbol{L_{align}}$ & \textbf{Tau} $\boldsymbol{\tau}$\\
    \midrule
    (a) & \checkmark & - & - & - & - & - & - & 0.5761  \\
 (b) & \checkmark & \checkmark & - & - & - & - & - & 0.5857 \up{0.96\%}   \\
(c) & \checkmark & - & \checkmark & - & - & - & - &0.5872 \up{1.11\%}\\
(d) & \checkmark & - & - & \checkmark & - & -  & - & 0.5256  \down{5.05\%}  \\
(e) & \checkmark & - & - & - & \checkmark & -  &  - &0.5824 \up{0.66\%} \\
(f) & \checkmark & - & \checkmark  & - & - & \checkmark  &  - & 0.5919 \up{1.58\%} \\
\cellcolor{Light}{(g)} & \cellcolor{Light}{\checkmark} & \cellcolor{Light}{-} & \cellcolor{Light}{\checkmark}  & \cellcolor{Light}{-} & \cellcolor{Light}{-} & \cellcolor{Light}{-}  &  \cellcolor{Light}{\checkmark} & \cellcolor{Light}{\textbf{0.5961} \up{2.00\%}} \\
% \midrule
% \multicolumn{9}{l}{\textit{PHD dataset}}  \\
% \midrule
% \textbf{Model6} & \checkmark & - & - & - & - & - & \multicolumn{1}{>{\centering\arraybackslash}p{2cm}}{-} & 21.75 \down{0.14\%}  \\
% \cellcolor{Light}{\textbf{\textbf{Ours}}} & \cellcolor{Light}{\checkmark} & \cellcolor{Light}{-} & \cellcolor{Light}{-} & \cellcolor{Light}{-} & \cellcolor{Light}{-} & \cellcolor{Light}{\checkmark} & \multicolumn{1}{>{\centering\arraybackslash}p{2cm}}{\cellcolor{Light}{-}} & \cellcolor{Light}{21.89}  \\
% \cellcolor{Light}{\textbf{ContentCTR-HL}} & \cellcolor{Light}{\checkmark} & \cellcolor{Light}{-} & \cellcolor{Light}{-}  & \cellcolor{Light}{-} & \cellcolor{Light}{-} & \cellcolor{Light}{\checkmark}  &  \cellcolor{Light}{-} &  \cellcolor{Light}{\textbf{21.89}   \\
\bottomrule
    \end{tabular}}
    \vspace{-8pt}
\end{table}

\begin{figure}[h]
\centering
\includegraphics[width=.47\textwidth]{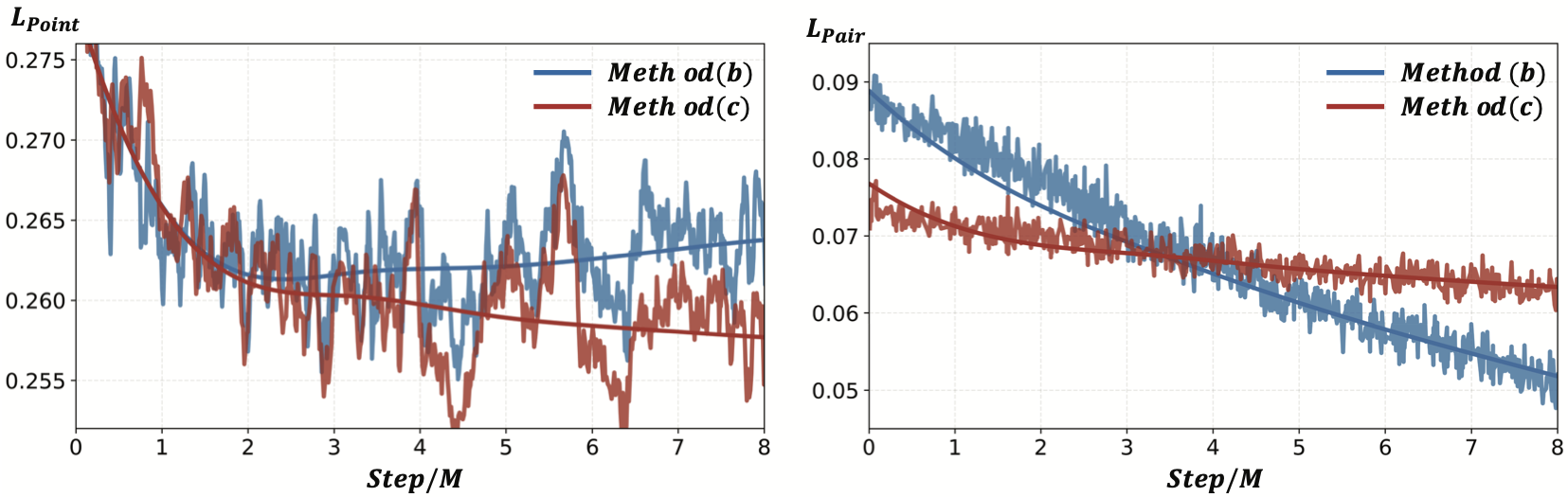}
\caption{\textbf{The change of $L_{Point}$ and $L_{Pair}$ during training.}}
\label{fig4}
\vspace{-0.2cm}
\end{figure}

When Method (b) optimizes $L_{Point}$ with unconstrained pairwise loss $L_{Pair}^{0}$, it shows a performance improvement of +0.96\%. However, we have noticed that during training, the pointwise loss of Method (b) tends to collapse due to the over-optimization of pairwise loss in Part 3, which is shown in Figure \ref{fig4}. So we investigate the performance of different versions of constrained pairwise loss with Method (c), (d) and (e). We find that Method (c) which is jointly optimized on Part 1 and Part 2 with $L_{Pair}^{1}$ shows a significant improvement of +1.11\% and both losses remain normal in Figure \ref{fig4}. Method (e) only shows a performance improvement of +0.66\%, while Method (d) with $L_{Pair}^{2}$ shows a significant performance degradation of -5.05\%. We hypothesize that the gradient changes are too drastic when optimizing only Part 1, which is detrimental to modeling the contrastive information of highlight frames and non-highlight frames. Therefore, we apply $L_{Pair}^{1}$ for final optimization.

%%%%%%%%%%%%%%%%%%%%%%%%%%%%%%%%%%%%%%
% According to Table \ref{lossfunc}, when we optimized Method (d) with $L_{Point}$ and $L_{Pair}^{2}$, its performance significantly degraded by -5.05\%. We hypothesize that the gradient changes are too drastic when optimizing only Part 1, which is detrimental to modeling the contrastive information of highlight frames and non-highlight frames.
% However, when optimizing only on Part 2 with $L_{Pair}^{3}$, Method (e) showed a performance improvement of +0.66\%, but it still falls behind the Method (c) which is jointly optimized on Part 1 and Part 2 with $L_{Pair}^{1}$, resulting in a significant improvement of +1.11\%.

% \begin{figure}[h]
% \centering
% \includegraphics[width=.49\textwidth]{fig4.pdf}
% \caption{\textbf{The change of $L_{Point}$ and $L_{Pair}$ during training.}}
% \label{fig4}
% \vspace{-0.2cm}
% \end{figure}

% Although Method (b) with $L_{Pair}^{0}$ has also shown performance improvement, we have noticed that during training, the pointwise loss of Method (b) tends to collapse due to the over-optimization of pairwise loss on Part 3 while both losses remain normal in Method (c), which is shown in Figure \ref{fig4}. 
%%%%%%%%%%%%%%%%%%%%%%%%%%%%%%%%%%%%%%

We also investigate the influence of MTAM alignment and our proposed augmentation method defined in Equation \ref{dtwdpdis1111}. When Method (f) incorporates $L_{align}^{'}$ for optimization, the naive DTW alignment loss achieves improvement with +1.58\%. By applying our proposed novel augmentation method to generate the positive sample, Method (g) gains further improvement by +0.42\%.

\begin{figure}[hbt]
% \vspace{-0.2cm}
\centering
\includegraphics[width=.47\textwidth]{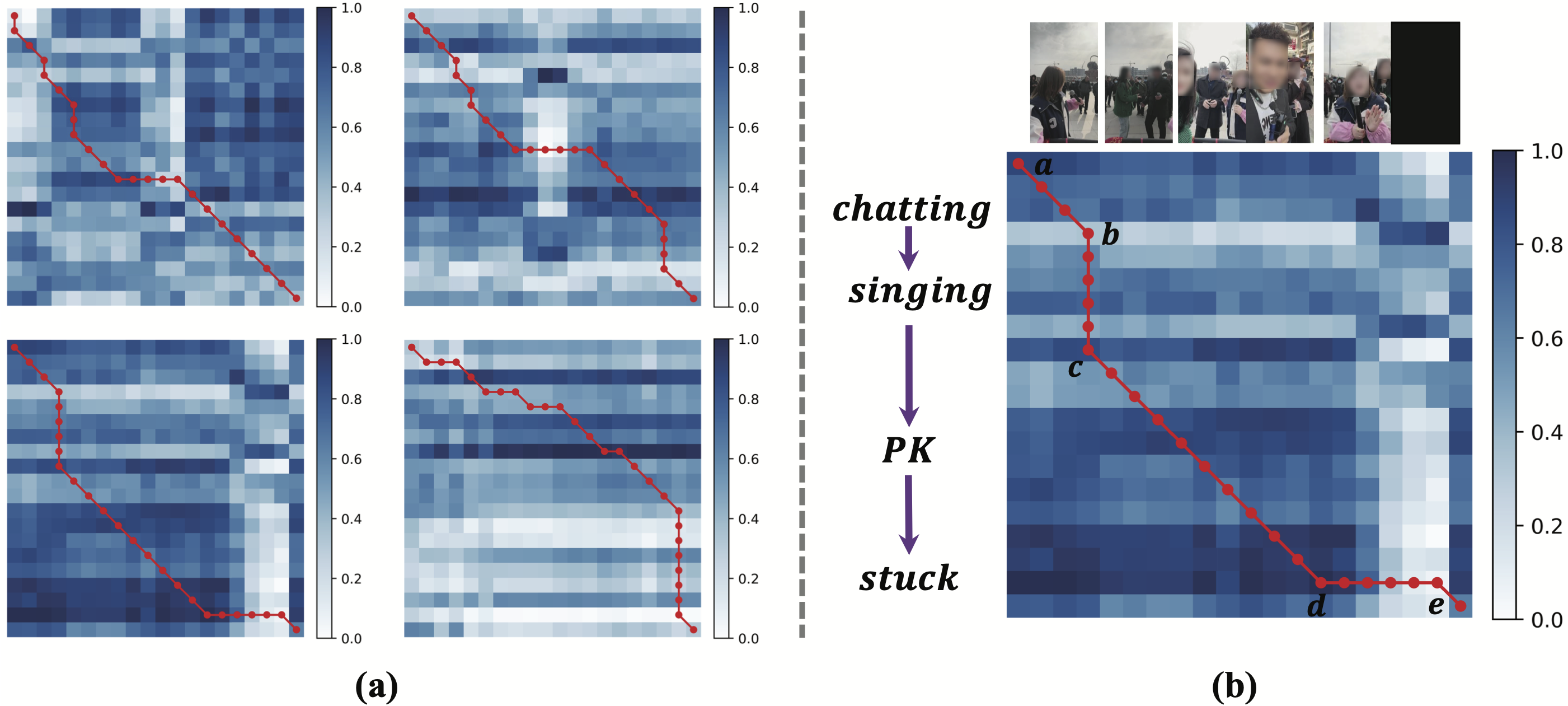}
\caption{\textbf{Visualization of Dynamic Time Warping (DTW) alignment results for the KLive dataset.}}
\label{fig5}
\vspace{-0.6cm}
\end{figure}

Figure \ref{fig5} (a) shows some visualizations of DTW alignment results for the KLive dataset, which reveal that misalignment cases do exist in streaming scenarios. Figure \ref{fig5} (b) presents a frame sequence and corresponding activities for one of the streaming scenarios. Here, the streamer shows her singing talent, where $a \rightarrow b$ indicates her interaction with the audience, resulting in aligned visual frames and text. However, during the transition from $b \rightarrow c$, the streamer starts singing, causing the text to become the lyrics of the song, resulting in a misalignment with the visual frames. As a result, the path becomes vertical. During $c \rightarrow d$, the streamer engages in a PK competition with another streamer. However, during $d \rightarrow e$, the frames from the other streamer get stuck due to network issues. Consequently, the visual frames become misaligned with text, causing the path to become horizontal. This case indicates that the involvement of $L_{align}$ does help our model in training better visual and text encoders that reduce the possible misalignment between the two.

\subsection{Modality Impact}
As shown in Table \ref{modallity}, on KLive we find visual modality has the most important impact, causing a performance degradation of -4.72\% when removed. The text modality is the second most significant factor, while the ID embedding has the smallest but still significant effect on the model's performance. On PHD dataset, the visual feature is more important than the captions modality.
% On PHD dataset, the visual feature of video frames is crucial for personalized video highlight detection, while the absence of captions resulted in a -1.11\% decrease in model's performance. 
\begin{table}[h]
    \centering
    \caption{
    Ablation study on different modality impact.
    }
    \label{modallity}
    % \vspace{-8pt}
    \setlength{\tabcolsep}{3.1pt}
    \scalebox{0.7}{
    \begin{tabular}{p{3.0cm}<{\raggedright}|p{0.5cm}<{\centering}p{0.5cm}<{\centering}p{0.5cm}<{\centering}p{0.5cm}<{\centering}p{0.5cm}<{\centering}|p{2.5cm}p{2.5cm}}
    \toprule
    Model & $v$ & $a$ & $x$ & $u$ & $c$ & \centering Tau $\tau$  & \multicolumn{1}{>{\centering\arraybackslash}p{2cm}}{mAP(\%)} \\
    \midrule
    \multicolumn{8}{l}{\textit{KLive dataset}}  \\
    \midrule
     \cellcolor{Light}{\textbf{KuaiHL}}  & \cellcolor{Light}{\checkmark} & \cellcolor{Light}{\checkmark} & \cellcolor{Light}{\checkmark} & \cellcolor{Light}{\checkmark} & \cellcolor{Light}{\textbf{-}} & \cellcolor{Light}{\textbf{0.5961}} & \multicolumn{1}{>{\centering\arraybackslash}p{2cm}}{\cellcolor{Light}{\textbf{-}}} \\
    \textbf{KuaiHL} w/o item & \checkmark & \checkmark & \checkmark & - & - &  0.5910 \down{0.51\%}&  \multicolumn{1}{>{\centering\arraybackslash}p{2cm}}{-} \\
     \textbf{KuaiHL} w/o text & \checkmark & - & - & \checkmark &  - & 0.5760 \down{2.01\%} &  \multicolumn{1}{>{\centering\arraybackslash}p{2cm}}{-}  \\
     \textbf{KuaiHL} w/o visual & - & \checkmark & \checkmark & \checkmark & - &  0.5489 \down{4.72\%} &  \multicolumn{1}{>{\centering\arraybackslash}p{2cm}}{-} \\
    \midrule
    \multicolumn{8}{l}{\textit{PHD dataset}}  \\
    \midrule
      \cellcolor{Light}{\textbf{KuaiHL}} &  \cellcolor{Light}{\checkmark} &  \cellcolor{Light}{-} &  \cellcolor{Light}{-} &  \cellcolor{Light}{-} &  \cellcolor{Light}{\checkmark} &  \multicolumn{1}{>{\centering\arraybackslash}p{2.5cm}}{\cellcolor{Light}{\centering -}}  &  \cellcolor{Light}{21.89}  \\
     \textbf{KuaiHL} w/o visual & - & - & - & - & \checkmark & \centering -  & 19.55 \down{2.34\%}  \\
     \textbf{KuaiHL} w/o caption & \checkmark & - & - & - & - & \centering -  & 20.06 \down{1.11\%}  \\
    \bottomrule
    \end{tabular}}
    \vspace{-15pt}
\end{table}

\section{Conclusion}
In this paper, we introduce a model called KuaiHL which utilizes a multimodal transformer to achieve frame-level LVTR prediction. Specifically, we propose a streamer-personalized Perceiver Block that fuses ID embedding, visual, audio, and comment embedding. The decoder network outputs the final LVTR prediction for each frame. To address the possible misalignment between video frames and texts, we carefully design a Modality Temporal Alignment Module for optimization. Additionally, the border-aware constrained pairwise loss demonstrates better performance when combined with the pointwise loss. We conduct comprehensive experiments on both the KLive dataset and the public PHD dataset, which demonstrate the effectiveness of our methods in both streaming LVTR prediction and video highlight detection tasks. Moreover, the proposed method has been deployed online on the company's short video platform and serves over three million daily users.

% \section*{References}
\section*{Acknowledgment}
This work was supported partially by Kuaishou through Kuaishou Research Intern Program, the National Natural Science Foundations of China (Grants No.62376267, 62076242) and the innoHK project.

% {\footnotesize
% \bibliographystyle{IEEEbib}
% \bibliography{icme2023template}
% }

\end{document}

\title{A Multimodal Transformer for Live Streaming Highlight Prediction Supplementary Materials\\
% {\footnotesize \textsuperscript{*}Note: Sub-titles are not captured in Xplore and
% should not be used}
}

\maketitle
\section{Details of Dataset}
\subsection{Details of KLive Dataset}

\textbf{Overview:} The live streaming highlight detection is first proposed by previous work \cite{zhao2022antpivot} and they also release a dataset called \textit{Anthighlight}. However, this dataset only contains 3,256 live streaming records and provides limited information, including the ASR (Automatic Speech Recognition) result of the streamer and human-annotated binary labels for highlight and no-highlight frames. Additionally, \textit{Anthighlight} primarily focuses on the topic of funds and wealth. So, the live streaming categories covered by \textit{Anthighlight} are also limited. In real-world scenarios, it may encounter the issue of human-annotated labels potentially invoking expert bias and being sub-optimal for reflecting the general preference of audiences. Thus, LSHD still lacks a large-scale multi-modal dataset that can serve as a benchmark. 

To address this gap, we construct the KLive dataset by collecting 17,897 live streaming records from the live streaming platform supported by KuaiShou. KLive contains nearly 19,334 hours of live streaming records across various streaming categories. The labels in our dataset represent the Long View Through Rate (LVTR) of corresponding live streaming clips, which are collected through user feedback. The major features of KLive include:
\begin{itemize}

  \item Open source: Our dataset provides a comprehensive benchmark in the field of live streaming highlight detection. Anyone can access the dataset with our company's approval.
  \item Multi-modality: Our dataset includes extracted features from various modalities in the live streaming scenario, including visual frames, interactive comments from viewers, and the ASR result of the hosts.
\item Dense annotation: Unlike traditional highlight detection datasets that only provide human-annotated binary labels, we provide the LVTR, which is analyzed from a large number of users' viewing histories. This metric better reflects users' general preferences towards streaming content.
\end{itemize}

\textbf{Data Collection and Labeling:} We select 17,897 high-quality live rooms based on a set of predefined criteria such as user rewards and viewing time duration. Each live room is then divided into multiple consecutive 30s live segments, with three pictures evenly sampled for each segment. The streamer's ASR and audiences' comments are extracted for each segment. We perform a preliminary filtering process on the obtained text results using regular expressions, such as removing emoticons and suppressing duplicate text and other noise content. Then, the frame images are resized to the shape of 3*224*224 and fed to a pre-trained Swin \cite{liu2021Swin}, resulting in representations with a size of 49*512. The ASR transcripts and comments text are truncated with a maximum token length of 200, and we extract the representations corresponding to the \texttt{<CLS>} token predicted by BERT \cite{devlin2018bert} as the textual embeddings. Next, we project the initial 768-dimensional features into 512-dimensional ones using a multi-layer perceptron. 

During the label-collecting stage, we expose the live frame segments obtained above to users and count the number of viewers for each live segment. We define users who watch for more than 60 seconds as positive samples. Therefore, the LVTR is calculated by dividing the number of positive users by all watched users. We then construct consecutive 20 streaming segments into a dataset sample, filtering out samples with fewer than 60 viewers of the most recent live segment to guarantee a reliable LVTR. We select 14,897 live rooms as the training set, while the remaining 3,000 live rooms as the test set. In the end, we obtained 1,436,979 and 286,510 samples for our training and test datasets, respectively. Each sample is comprised of the streamer's ID, three video frames visual features for 20 streaming segments, audience comments and streamer's ASR speech. Because the objective of our method is to predict the highlight moment of future frames, we left shift the ground truth LVTR for all live segments as the final label

\begin{figure}[h]
\centering
\includegraphics[width=.49\textwidth]{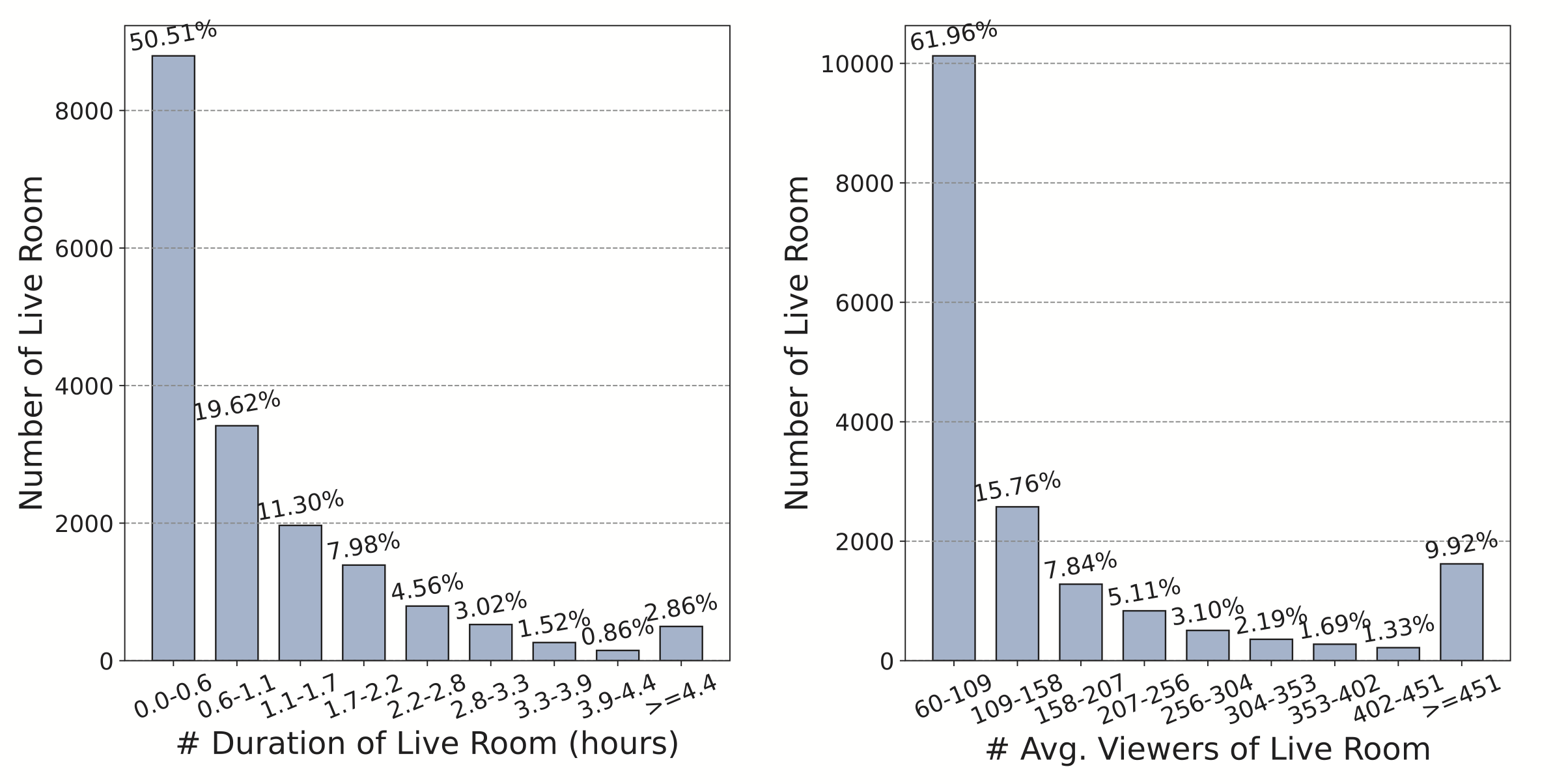}
\caption{\textbf{Distribution of number of live room w.r.t. the duration and average viewers number.}}
\label{fig7}
% \vspace{-0.7cm}
\end{figure}

\textbf{Statistical Analysis:} In KLive, the total number of live rooms is 17,897, which comes from 9,932 hosts and has a total of almost 19,334 hours of live streaming records. The longest recorded live streaming room lasts for 46.46 hours, while the average duration is 1 hour. The maximum average number of viewers in these live rooms is over 32,006, and the minimum is 60 (as we have filtered out live rooms with less than 60 viewers). We visualize the distribution of the number of live rooms based on their duration and average number of viewers in Figure \ref{fig7}. From the figure, we can see that 49\% of live rooms in this dataset have durations greater than 36 minutes, and 38\% of live rooms have an average viewership of over 100. 

\textbf{Release Agreement:} 
One must sign the following agreement in order to be granted permission to use our dataset.

\begin{framed}
\begin{center}
\textbf{\textit{KLive} Dataset Usage Agreement}
\end{center}
% \begin{itemize}

$\bullet$ Non-Commercial Use Only: The usage of our dataset is restricted to non-commercial purposes, such as academic research and education. Any activities that aim to generate profit or infringe upon others' rights, such as advertising, selling, or developing facial recognition applications, are strictly prohibited.

$\bullet$ No Analysis of Bio-Info: You agree not to perform any analysis related to bio-information in the dataset, including but not limited to facial features, gender, and so on.

$\bullet$ Responsibility for Usage: You acknowledge that you assume full responsibility for your usage of the dataset. If you engage in any illegal activities or cause any negative consequences, we reserve the right to request an immediate cessation of usage and the elimination of any associated impact.

$\bullet$ Right to Terminate: In the event of any violation of this agreement, we retain the right to terminate your access to the dataset and delete any copies in your possession.

$\bullet$ Purpose and Potential Effects: You must clearly state the purpose and potential effects of utilizing our dataset.
% \end{itemize}
\begin{center}
\textbf{Name:\underline{\quad}}\textbf{Email:\underline{\quad}}\textbf{Organization:\underline{\quad}}\textbf{Date:\underline{\quad}}\textbf{Signature:\underline{\quad}}
\end{center}

\end{framed}
\subsection{Details of PHD Dataset}
To verify the generality of our method, we evaluate the publicly available personalized video highlight detection dataset \cite{garcia2018phd} (PHD). This dataset comprises of URLs of YouTube videos, IDs of evaluators or "users," and the segments they designated as highlight frames based on their preferences. The most recent video that a user annotated is considered as the target video for that particular user. Since PHD only provides YouTube URLs, we download the original videos and crawl the captions from YouTube to carry out the experiments. Our training and testing sets include up to 20 history highlight frames and one target video per user. The duration of the history highlight frames varies from 1 to 672.19 seconds, with an average length of 5.12 seconds. The target videos range from 1 to 37,434 seconds, with an average length of 431.79 seconds. The training set comprises a total of 12,541 users and 81,056 videos, while the testing set has 833 users and 7,595 videos that do not overlap with any user or video in the training set. Consistent with \cite{chen2021pr}, we divide the target video into fixed-length segments (192 frames) and only train those segments that contain highlight frames. During testing, we make the entire video segments as input for inference.

\section{Implementation Details}
\textbf{Baseline:} On KLive dataset, we compare our method with livestream highlight detection model and several strong baselines of personalized video highlight detection as follows:
\begin{itemize}[leftmargin=*]
\item \textbf{AntPivot} \cite{zhao2022antpivot} leverages a hierarchical attention module to model the highlight detection task in a dynamic programming manner.
\item \textbf{Adaptive-H-FCSN} \cite{rochan2020adaptive} employs a convolutional highlight detection network with a history encoder to learn user-specific highlight patterns.
\item \textbf{PR-Net} \cite{chen2021pr} proposes a reasoning framework to explicitly learn frame-level patterns. It also employs contrastive learning to alleviate annotation ambiguity.
\item \textbf{PAC-Net} \cite{wang2022pac} introduces a Decision Boundary Customizer (DBC) module and a Mini-History (Mi-Hi) mechanism to capture more fine-grained user-specific preferences.
\item \textbf{ShowMe} \cite{bhattacharya2022show} leverages the content of both user history and target videos, using pre-trained features of YOLOv5 \cite{Jocher_YOLOv5_by_Ultralytics_2020}.
\end{itemize}
On the PHD dataset, to compare our method with the aforementioned several strong baselines for personalized video highlight detection, we have made slight modifications to our proposed methods. These modifications are as follows: First, for a fair comparison, we extract YOLOv5 features from images for training and testing, which is also commonly adopted by previous works. Second, the ID embedding is replaced with the user history feature for personalized highlighting. The text comes from the captions crawled from the YouTube video. Then, since there is no need to consider future information leakage in the video highlighting task, the causal attention mask is set to fully visible. Finally, when optimizing the network, only $\mathcal{L}_{Point}$ and $\mathcal{L}_{align}$ are considered for optimization. This is because the PHD dataset only provides binary labels of either highlight (1) or no-highlight (0) for each frame, so $\mathcal{L}_{Pair}$ is meaningless in this task.

\textbf{Setup Detail:} During the training of our model on the KLive dataset, $L$ of Pereciver block is set to 3, the input dimension $d=512$, the hidden dimension of the transformer $d_h$ is set to 64 and the number of attention head $n_h=8$. We utilize pre-trained swin as the Encoder $F(\cdot)$ and pre-trained Bert Chinese Large as the Encoder $G(\cdot)$ and they are both pre-trained on image and text data from 17,897 streaming rooms. The tradeoff parameters $\lambda_{1}=0.65$, $\lambda_{2}=0.15$, and $\lambda_{3}=0.20$. The $\sigma$ in $L_{Pair}^{1}$ is set to 10 and the number of negative sample $N$ is set to 8. We optimize our model for 12 epochs with a learning rate of $5 \times 10^{-5}$ using the Adam optimizer, and the global batch size is set to 48. During the training and testing of ID-based recommendation methods on the KLive dataset, the input sparse features for these methods include the streamer item ID, live ID, timestamp ID, exposure count, comment count, gift count, click count, like count, follow count, room entry and exit count for each streaming segment. The objective of these methods is to regress the LVTR of each segment. 

When training our model on the PHD dataset, we first extract the YOLOv5 features for each frame of the historical highlight frames in the training and testing set. We use pre-trained YOLOv5 as the Encoder $F(\cdot)$ and pre-trained Bert English Large as the Encoder $G(\cdot)$. Then we train our network using the Adam optimizer with a batch size of 8 and an initial learning rate of $5 \times 10^{-5}$ for 20 epochs.

\section{More Experimental Results}

\subsection{Online Experiments}
\begin{spacing}{1.0}
We test the proposed framework in real-world live streaming scenarios through online A/B testing. The experiment is conducted over four consecutive days, with traffic randomly assigned to either the baseline method or our method. In the baseline group, candidate live rooms are sorted by scores produced by a traditional recommendation model, e.g., Click-Through Rate $s_{ctr}$, Long-View-Through Rate $s_{lvtr}$. Note that these scores focus on capturing the long-term relationship between streamers and users. In contrast, our method utilizes content-based CTR as an additional factor like $s_{exp}$. This term is capable of catching the highlight moment and showing the most attractive live content to users. The results show that our method achieve 2.9\% and 5.9\% improvements in terms of CTR and live play duration, respectively, which demonstrate the effectiveness of our content-based model.
\end{spacing}

{\footnotesize
\begin{equation}
\begin{aligned}
s_{base} & = s_{ctr} + s_{lvtr} + ... \\
s_{exp} & = s_{ctr} + s_{lvtr} + ... + s_{KuaiHL},
\end{aligned}
\end{equation}
\label{abtest}
}

% {\footnotesize
% \bibliographystyle{IEEEbib}
% \bibliography{icme2023template}
% }